\documentclass[pdflatex,sn-mathphys-num,a4paper]{sn-jnl}

\usepackage{physics}

\usepackage{graphicx}%
\usepackage{multirow}%
\usepackage{amsmath,amssymb,amsfonts}%
\usepackage{amsthm}%
\usepackage{mathrsfs}%
\usepackage[title]{appendix}%
\usepackage{xcolor}%
\usepackage{textcomp}%
\usepackage{manyfoot}%
\usepackage{booktabs}%
\usepackage{algorithm}%
\usepackage{algorithmicx}%
\usepackage{algpseudocode}%
\usepackage{listings}%

\theoremstyle{thmstyleone}%
\theoremstyle{thmstyletwo}%

\theoremstyle{thmstylethree}%

\begin{document}

\title[Article Title]{Fragmentation temperature of 1D and 3D quantum droplets in a BEC mixture}


\author*[1]{\fnm{Jeroen} \sur{Van Loock}}\email{jeroen.vanloock@uantwerpen.be}

\author[1]{\fnm{Denise} \sur{Ahmed-Braun}}

\author[1]{\fnm{Jacques} \sur{Tempere}}

\affil[1]{\orgdiv{TQC}, \orgname{Universiteit Antwerpen}, \orgaddress{\street{Universiteitsplein 1}, \city{Antwerp}, \postcode{2610}, \country{Belgium}}}


\abstract{
In a mixture of two Bose-Einstein condensates, the interactions can be tuned such that self bound objects called quantum droplets appear. Whereas the ground states of such quantum droplets at finite temperature have been studied for three- and one-dimensional configurations, the possible fragmentation of these droplets has so far not been considered in these studies. 
In this paper we show that droplets can lower their free energy by splitting or fragmenting in a combination of multiple smaller droplets and/or a gas. 
Three-dimensional droplets will split when the interspecies interaction strength is considerably stronger than the intraspecies interaction strength, and the number of atoms is of the same order as the minimum number of atoms necessary to 
form a droplet. 
One-dimensional droplets will fragment as long as the intraspecies and interspecies interactions strength do not vary too much in strength and the density is not to big compared with the scattering length. If the temperature rises, 1D droplets will split by expelling atoms, forming a gas of predominantly free atoms and pairs of atoms. 
These pairs remain present in the system up to considerably high temperatures compared to the transition temperature. 
Our results provide important insights on the stability of these droplets.

}




\maketitle

\section{Introduction}\label{sec: introduction}
Ultracold atomic gases provide a highly controllable platform for exploring quantum many-body phenomena. 
In three-dimensional (3D) systems with attractive interaction, the mean-field theory predicts collapse \cite{pitaevskij2016}. However, in mixtures of Bose-Einstein condensates (BECs) where the attractive interspecies interactions are almost canceled by the repulsive intraspecies interactions, quantum fluctuations beyond mean-field will stabilize the system, leading to the formation of \textit{self-bound quantum droplets} in free space. 
These droplets are dilute, liquid-like states whose stability arises from the Lee-Huang-Yang (LHY) correction \cite{lee1957} to the mean-field energy, which introduces a repulsive term scaling as $n^{5/2}$ that counteracts the attractive mean-field term scaling as $n^2$ \cite{petrov2015} that would normally lead to a collapse of the gas. In one-dimensional systems, the LHY energy is attractive and scales as $n^{3/2}$. Quantum droplets will thus form when the mean field energy, which still scales as $n^2$, is repulsive \cite{petrov2016}. Note that in dipolar gasses, quantum droplets that are stabilized by the same principle also exist \cite{ferrier-barbut2016a, chomaz2016,ferrier-barbut2016,kadau2016}. \\

Quantum droplets in 3D systems have been experimentally realized in homonuclear and heteronuclear Bose mixtures, notably in potassium-39 \cite{cabrera2018a, semeghini2018, ferioli2019a} and potassium-rubidium mixtures \cite{derrico2019, cavicchioli2025,cheiney2018}. These experiments confirm the existence of droplets stabilized solely by contact interactions, and reveal phenomena such as droplet fission driven by capillary instability \cite{cavicchioli2025, ancilotto2023, ancilotto2025}. Theoretical studies have further explored the ground-state properties and phase behavior of these systems \cite{pan2022, flynn2023a}, including the role of imbalance in atomic populations and the effects of higher-order quantum and thermal fluctuations. For 1D droplets, there has recently been a surge in theoretical effort to describe droplet properties, such as splitting and fragmentation in harmonic traps
\cite{pathak2022,bristy2025}. Droplets can also form in 2D systems \cite{petrov2016}, and could be treated in the same way as the 3D and 1D systems in this paper. However, we will limit this study to the 3D and 1D systems. \\

While the existence (in 3D) and the stability of quantum droplets are well established, their thermodynamic behavior at non-zero temperature remains less understood, and has so far only been studied in the context of single droplets \cite{guebli2021,ota2020,wang2020a,boudjemaa2023a}. 
However, the nonlinear scaling of energy with particle number introduces a competition between energy and entropy: a single droplet may be energetically favorable, but fragmentation into smaller droplets or individual atoms can be entropically preferred. This competition suggests the possibility of a \textit{fragmentation transition}, analogous to a vaporization process, where the system transitions from a single bound state droplet to a fragmented configuration \cite{atkins2023}. A similar fragmentation transition \cite{herzog2014} has been predicted for solitons in attractive Bose gases, in a one-dimensional system where bright solitons are stabilized as the collapse from the attractive interactions is balanced by the kinetic energy. Since there exists a smooth connected phase diagram that connects these bright solitons to quantum droplets  \cite{cheiney2018}, this implies that  that fragmentation could also potentially be observed in quantum droplets. \\

To investigate this transition, the free energy of the system is computed as a function of temperature, atom number, and interaction parameters \cite{herzog2014}. Comparing configurations ranging from a single droplet to maximally fragmented states enables the construction of a phase diagram that characterizes the conditions under which fragmentation occurs. 
In this work, we focus on the equilibrium thermodynamic properties of quantum droplets, rather than the dynamics of the fragmentation process, in both three-dimensional and one-dimensional Bose mixtures. Our goal is to characterize the fragmentation transition and map out the phase diagram for fragmentation, thereby contributing to a deeper understanding of self-bound quantum matter and the role of many-body effects in determining equilibrium configurations. 
Our paper is structured as follows: in section \ref{sec: theory}, the ground state of single 3D and 1D quantum droplets is derived. In section \ref{sec:Theory of droplet fission}, the theoretical framework is set up to find the ground state of multiple quantum droplets. In section \ref{sec: results}, the results are shown and discussed and finally in section \ref{sec concl outlook} the work is concluded.

\section{Theory of quantum droplets}\label{sec: theory}
In this work, we consider a two-component bosonic gas in both three dimensions and one dimension, described by the following Hamiltonian \cite{pitaevskij2016, larsen1963}
\begin{align}\label{eqn: full ham two comp}
   \hat{H}=&\sum_{j=1}^{2}\left[ \int d\vec{r} \hat{\psi}_{j}^{\dagger}(\vec{r})h_{j}\hat{\psi}_{j}(\vec{r})\right] 
   + \frac{g_{jj}}{2}\int d\vec{r}\hat{\psi}_{j}^{\dagger}(\vec{r})\hat{\psi}_{j}^{\dagger}(\vec{r})\hat{\psi}_{j}(\vec{r})\hat{\psi}_{j}(\vec{r})\Big] \\
   &+g_{12}\int d\vec{r}  \hat{\psi}_{1}^{\dagger}(\vec{r})\hat{\psi}_{2}^{\dagger}(\vec{r}^{\prime})\hat{\psi}_{2}(\vec{r}^{\prime})\hat{\psi}_{1}(\vec{r}).
 \end{align}
Here, the field operators $\hat{\psi}_{j}^{(\dagger)}(\vec{r})$ annihilate (create) particles of species $j$ at position $\vec{r}$. Furthermore, the single-particle Hamiltonian $h_j$ for each species is defined as
 \begin{equation}
   h_j = -\frac{\hbar^2}{2m_j} \laplacian + U_j(\vec{r}) - \mu_j,
 \end{equation}
 where $m_j$ denotes the mass of species $j$, $U_j(\vec{r})$ represents an external trapping potential and where $\mu_j$ represents the chemical potential. Considering only the s-wave interactions that dominate in the ultracold regime, in Eq.~\eqref{eqn: full ham two comp} both the interspecies interactions and intraspecies interactions are modeled using contact potentials, and are linearly related to the 3D, or 1D, scattering lengths $a$ and $a^{\mathrm{1D}}$ respectively. \\
 
 Specifically, the 3D interspecies interaction strengths $g_{jj}$ and the intraspecies interaction strength $g_{ij}$ correspond to \cite{pitaevskij2016} 
 \begin{equation}
g_{jj} = \frac{2\pi \hbar^2 a_{jj}}{m_{jj}} \qq{and} g_{ij} = \frac{2\pi \hbar^2 a_{ij}}{m_{ij}},
 \end{equation}
 where the reduced masses $m_{ij} = \frac{m_i m_j}{m_i+m_j}$ is introduced. 
However, direct implementation of 3D delta-function potentials with strengths $g_{ij}$ and $g_{jj}$ is problematic, since contact interactions in 3D introduce UV divergences in many-body calculations, particularly in perturbative expansions and when computing the T-matrix \cite{braaten2006}. To address these divergences, the standard procedure of regularization and renormalization can be applied. To this extent, we introduce a momentum-space cut-off $\Lambda$ and define the following renormalization relation \cite{wang2022,hu2020b,braaten2006}
\begin{align} \label{eqn: renormalization MF}
\frac{1}{\bar{g}_{ij}} = \frac{1}{g_{ij}}-\sum_{\vec{k}}^{\Lambda} \frac{2 m_{ij}}{\hbar^2 k^2},
\end{align}
with the renormalized interaction strengths $\bar{g}_{ij}$ and the bare interaction strengths $g_{ij}$.
This renormalization ensures that physical quantities such as energy and density remain well-defined and independent of the regularization scheme. \\

Contrary to the 3D contact interactions, the 1D contact interactions do not require renormalization, since no divergences in 1D momentum-space integrals arise on the level of the two-body scattering amplitude. As such, we use the following interspecies interaction strength $g_{ij}$ and intraspecies interaction strength $g_{jj}$ directly in our subsequent 1D analyses \cite{levin2012,ota2020,parisi2019,petrov2023}
\begin{align}\label{eqn gij 1d}
g_{ij}^{\mathrm{1D}} = -\frac{\hbar^2}{a^{\mathrm{1D}}_{ij}m_{ij}} \qq{and} g_{jj}^{\mathrm{1D}} = -\frac{\hbar^2}{a^{\mathrm{1D}}_{jj}m_{jj}} .
\end{align}
The superscript in $g^{\mathrm{1D}}$ and $a^{\mathrm{1D}}$, will be omitted for the remainder of the paper.

\subsection{Beyond Mean-field analysis of BEC mixtures}
Having set the general framework for our analysis in the previous section, we now proceed to analyze the stabilization criteria for quantum droplets in 3D and 1D. As previously mentioned in the introduction, for BEC mixtures, these droplets form in a precarious regime where the first beyond mean-field correction, the so-called LHY correction, stabilizes the gases that are either very close to the expanding regime (for 3D quantum droplets), or very close to the regime of collapse (for 1D quantum droplets). In order to obtain the ground state order parameters for these droplets, we aim to derive the extended Gross-Pitaevskii equation (eGPE) for the BEC mixture, which we will then solve by using imaginary time propagation of the ground-state solutions. The first step in this derivation amounts to computing the mean-field and LHY energy densities which can be used to derive the eGPE equation.  
\subsubsection*{Mean-field and LHY energy densities}
In order to find the mean-field and LHY energy densities, the field operators $\hat{\psi}_j$ in the Hamiltonian as presented in Eq.~\eqref{eqn: full ham two comp} are replaced with 
\begin{align} \label{eqn: two comp eta}
\hat{\psi}_j(\vec{r},t) = (\psi_j(\vec{r}) + \hat{\eta}_j(\vec{r},t))e^{-i\mu_j t/\hbar},
\end{align}
where $\psi_j(\vec{r})$ corresponds to the classical mean-field value defined by the expectation value $\psi_j = \ev{\hat{\psi}_j}$, and where $\hat{\eta}_j(\vec{r},t)$ corresponds to the quantum fluctuations of the field $\hat{\psi}_j(\vec{r},t)$ that will generate the LHY correction to the mean-field energy \cite{petrov2023,pitaevskij2016,fetter2012}. 
Substituting Eq.~\eqref{eqn: two comp eta} into Eq.~\eqref{eqn: full ham two comp} and keeping only fluctuation corrections up to quadratic order (corresponding to the so-called Bogoliubov approximation), results in
\begin{align}
\label{eq:Hsplit}
    \hat{H} \approx H_0  + \hat{H}_2.
\end{align}
Here, $H_0$ corresponds to the mean-field energy, whereas the expectation value of $H_2$ yields the LHY energy density. We will now proceed to discuss these contributions separately. \par\vspace{1em}

\noindent\textbf{Mean-field energy density} \\
Focusing first on the mean-field energy given by $H_0$, results in
\cite{pitaevskij2016}
\begin{equation}\label{eqn: full E two comp}
   E_\mathit{MF} = \int \qty[ \sum_{j=1}^{2}\left[  \psi_{j}^{*}(\vec{r})\qty(-\frac{\hbar^2}{2m_j} \laplacian + U_j(\vec{r}))\psi_{j}(\vec{r})
   + \frac{g_{jj}}{2} n_j^2(\vec{r})\right] 
   +g_{12}  n_{1}(\vec{r}) n_2(\vec{r})]d\vec{r},
 \end{equation}
where $n_j=\psi_j^*\psi_j$ represents the density of the components. Under the Thomas-Fermi approximation and in the absence of external trapping potentials, Eq.~\eqref{eqn: full E two comp} allows for two stable BEC solutions: one where the two components mix, with associated energy $E_{mix}$ and one where the two components remain separated, with associated energy $E_{sep}$, given respectively by \cite{pitaevskij2016} 
\begin{equation}\label{eqn: mix basic energie}
   E_\mathit{mix} = \frac{g_{11}}{2} \frac{N_1^2}{V} + \frac{g_{22}}{2} \frac{N_2^2}{V} + g_{12} \frac{N_1N_2}{V}
\end{equation}
and
\begin{equation}
E_\mathit{sep} = \frac{g_{11}}{2} \frac{N_1^2}{V} + \frac{g_{22}}{2} \frac{N_2^2}{V} + \sqrt{g_{11}g_{22}}\frac{N_1N_2}{V}.
\end{equation}
Where $N_1$ and $N_2$ are the number of atoms in each component. Evidently, the equilibrium ground state of the system will be determined by the solution with the lowest energy, such that the comparison of the previous two expressions sets the condition that: if $g_{12} > \sqrt{g_{11}g_{22}}$ the system will separate and if $g_{12} < \sqrt{g_{11}g_{22}}$ the system will mix. Here, we note that if $g_{11}$ and $g_{22}$ are repulsive, the presence of stable solutions rests on the assumption that the absolute value of $g_{12}$ is not larger in magnitude than $\sqrt{g_{11}g_{22}}$, since in this case the BECs will collapse altogether. Aiming to study quantum droplets, which will occur around the collapse/expansion transition point, it can be assumed that $\delta g= g_{12}+\sqrt{g_{11}g_{22}}$ is small. Then, performing a Taylor expansion of the mean-field energy $E_{MF}$ around $\delta g \rightarrow 0$ and keeping only terms up to linear order in $\delta g$, results in \cite{petrov2015}
\begin{align}\label{eqn: def energy density}
   E_\mathit{MF} = \int \epsilon_\mathit{MF} d\vec{r} \qq{with}
    \epsilon_\mathit{MF} = \sum_{i,j} \frac{g_{ij}}{2} n_i n_j = \sum_{\pm}\lambda_\pm n_\pm^2,
\end{align}
where $\epsilon_{MF}$ represents the mean-field energy density.
In the last step of the previous equation, the expression for the energy density $\epsilon_{MF}$ has been diagonalized, with eigenvalues $\lambda_\pm$ and eigenvectors $n_{\pm}$ corresponding respectively to
 \begin{align}
      \lambda_+ \approx \frac{g_{11}+g_{22}}{2} \qq{and} \lambda_- \approx \frac{\sqrt{g_{11}g_{22} }\delta g}{g_{11}+g_{22}}, 
 \end{align}
and
\begin{align}
   n_+ \approx \frac{\sqrt{g_{11}}n_1 - \sqrt{g_{22}} n_2}{\sqrt{g_{11}+g_{22}}} \qq{and}
   n_- \approx \frac{\sqrt{g_{22}}n_1 + \sqrt{g_{11}} n_2}{\sqrt{g_{11}+g_{22}}},
 \end{align}
for small values of $\delta g$. Where $n_+$ is related to the total density, and $n_-$ is related to the difference in density between the two components. 
Since $\lambda_+$ is much larger than $\lambda_-$, the total energy is minimized when the two densities have the same profile, which is equivalent to minimizing $n_+$ and leads to
 \begin{align}\label{eqn: def gamma}
   n_2 = \sqrt\frac{{g_{11}}}{{g_{22}}} n_1 \equiv \gamma n_1.
 \end{align}
 A more rigorous derivation for this ratio ($\gamma = \sqrt{g_{11}/g_{22}}$) can be found in \cite{petrov2023}. If the densities are not in this equilibrium, the majority component will first be absorbed within the droplet until it saturates, then the excess atoms will be expelled out of the droplet \cite{flynn2023a}. In this work, it is assumed that the densities are in equilibrium. The energy density (Eq. (\ref{eqn: def energy density})) can then be written as \cite{petrov2015}
 \begin{align}\label{eqn: mean field energy}
   \epsilon_\mathit{MF} \approx \frac{\sqrt{g_{11}g_{22} }\delta g}{g_{11}+g_{22}} \qty(\frac{\sqrt{g_{22}}n_1 + \sqrt{g_{11}} n_2}{\sqrt{g_{11}+g_{22}}})^2 =  \delta g \sqrt{\frac{{g_{11}}}{{g_{22}}}} n_1^2 =\frac{\delta g \gamma}{(1+\gamma)^2}n^2,
 \end{align}
 with $n=n_1+n_2$. Equation~\eqref{eqn: mean field energy} will be used in our derivation of the eGPE as presented in Sec.~\ref{sec eGPE}. \par\vspace{1em}
\noindent\textbf{LHY energy density} \\
Having obtain the mean-field energy density $\epsilon_{MF}$ in the previous subsection, we now proceed to compute the LHY contribution to the energy density. To this extent, we now focus on the term $\hat{H}_2$ in Eq.~\eqref{eq:Hsplit}. Here, $\hat{H}_2$ represents the piece of the Hamiltonian scaling with quadratic corrections in $\hat{\eta}_j(\vec{r},t)$. In momentum space, using the momentum-space operator defined as $\hat{\eta}_j(\vec{r},t) = \sum_{\vec{k} \neq 0} \hat{a}_{j,\vec{k}}(t) e^{i\vec{k} \cdot \vec{r}}$, the contribution $\hat{H}_2$ can be written as \cite{petrov2023}
\begin{align}\label{eqn: H2 non diag}
   \hat{H}_2=-\frac{1}{2}\sum_{j,\vec{k}\neq 0}\left[\frac{\hbar^2k^2}{2m_j}+g_{jj}n_j\right]+\frac{1}{2}\sum_{\vec{k}\neq 0}(\hat{a}_{1,\vec{k}}^\dagger\hat{a}_{1,-\vec{k}}\hat{a}_{2,\vec{k}}^\dagger\hat{a}_{2,-\vec{k}})h_2
   \begin{pmatrix}
   \hat{a}_{1,\vec{k}} \\
   \hat{a}_{1,-\vec{k}}^\dagger \\
   \hat{a}_{2,\vec{k}} \\
   \hat{a}_{2,-\vec{k}}^\dagger
   \end{pmatrix}
 \end{align}
 with
 \begin{align}
    h_2 = \begin{pmatrix}
     \frac{\hbar^2k^2}{2m_1} + g_{11} \abs{\psi_1}^2  & g_{11} (\psi_1)^2 & g_{12} \psi_1 \psi_2^* & g_{12}\psi_1\psi_2\\
     g_{11}(\psi_1^*)^2 & \frac{\hbar^2k^2}{2m_1} + g_{11}\abs{\psi_1}^2  & g_{12} \psi_1^*\psi_2^* & g_{12} \psi_1^*\psi_2\\
     g_{12} \psi_1^*\psi_2 & g_{12}\psi_1\psi_2 & \frac{\hbar^2k^2}{2m_2} + g_{22}\abs{\psi_2}^2 & g_{22} (\psi_2)^2\\
     g_{12}\psi_1^*\psi_2^* & g_{12} \psi_1\psi_2^* & g_{22} (\psi_2^*)^2 & \frac{\hbar^2k^2}{2m_2} + g_{22}\abs{\psi_2}^2
   \end{pmatrix}.
 \end{align}
Now, in order to find the fluctuation energy,  Eq.~\eqref{eqn: H2 non diag} is diagonalized, resulting in \cite{petrov2023,ota2020,ota2020a}
\begin{align}\label{eqn: H2 first}
   \hat{H}_2=-\frac{1}{2}\sum_{j,\vec{k}\neq 0}\left[\frac{\hbar^2k^2}{2m_j}+g_{jj}n_j\right]+\frac{1}{2}\sum_{i= \pm,\vec{k}\neq 0} E_{i} \qty(\hat{\alpha}_{i,\vec{k}}^\dagger \hat{\alpha}_{i,\vec{k}} + \hat{\alpha}_{i,\vec{k}} \hat{\alpha}_{i,\vec{k}}^\dagger),
 \end{align}
where the quasiparticle fields $\alpha_{+,\vec{k}}$ and $\alpha_{-,\vec{k}}$ are introduced and where $E_{\pm}$ correspond to eigenenergies \cite{petrov2023,alexandrov2002,ota2020,ota2020a}
 \begin{align}\label{eqn: double energietakken}
   E_{\pm}(\vec{k})=2^{-1/2}\left(\epsilon_1^2(\vec{k})+\epsilon_2^2(\vec{k})\pm\sqrt{[\epsilon_1^2(\vec{k})-\epsilon_2^2(\vec{k})]^2+\frac{4\hbar^4k^4}{m_1m_2}g_{12}^2n_1n_2}\right)^{1/2},
 \end{align}
with  Bogoliubov energies $\epsilon_j$ given by 
 \begin{align}
     \epsilon_{j}(\vec{k})=\sqrt{\hbar^4k^{4}/(4m_{j}^{2})+\hbar^2 k^{2}g_{jj}n_{j}/m_{j}}.
 \end{align}
In the absence of Bogoliubov quasiparticle excitations, formally corresponding to the zero-temperature limit, we find the LHY energy density by computing the expectation value of Eq.~\eqref{eqn: H2 first}, resulting in \cite{petrov2023,ota2020}
 \begin{align}\label{eqn: LHY met g gewoon}
   \epsilon_\mathit{LHY} = \frac{1}{2}\sum_{j,\vec{k}\neq 0}\left[(E_+ +E_-) - \qty(\frac{\hbar^2 k^2}{2m_j}+g_{jj}n_j)\right].
 \end{align}
What remains now is to perform the summations over the momenta present in Eq.~\eqref{eqn: LHY met g gewoon} to obtain explicit solutions to the LHY energy, with which we shall now proceed.
\par\vspace{1em}
\noindent\textbf{Explicit results for the 3D and 1D LHY energy densities}\\
As pointed out at the beginning of Sec.~\ref{sec: theory}, performing summations over momenta becomes problematic for the 3D analysis, where the use of the contact interaction results in divergences in the momentum-space integral. 
Examination of Eq.~\eqref{eqn: LHY met g gewoon} reveals that the UV divergence in 3D can be exactly canceled in the LHY energy density by adding an additional term. To keep the total energy density unaltered, the same term has to be subtracted from the MF term in Eq. (\ref{eqn: mean field energy}). The LHY energy density will thus be redefined to \cite{petrov2015,petrov2023} 
 \begin{align}\label{eqn: two comp lhy with sum over k}
   \epsilon_\mathit{LHY}^{3D}  = \frac{1}{2} \sum_{\vec{k}\neq 0}\qty[\sum_j E_j - \qty(\frac{\hbar^2 k^2}{2m_j}+g_{jj}n_j) + \sum_i  \frac{2 m_{ij} g_{ij}^2 n_in_j}{\hbar^2 k^2}],
 \end{align}
and the mean-field energy density to
 \begin{align}\label{eqn: mean field niewe g}
   \epsilon_\mathit{MF}^{3D} = \frac{1}{2} \sum_{i,j} n_i n_j \qty[g_{ij} - \sum_{k\neq 0} \frac{2 m_{ij} g_{ij}^2}{\hbar^2 k^2}].
 \end{align}
To cure the divergence that now occurs in the expression for $\epsilon_\mathit{MF}^{3D}$, the summation has to be regulated by introducing the cut-off $\Lambda$ (where $\Lambda \gg 1$), and renormalized by replacing the interaction strength with the renormalized interaction strengths $\bar{g}_{ij}$ as defined in Eq.~\eqref{eqn: renormalization MF}. Then, by taking the continuum limit and explicitly performing the integration, our expression for the mean-field energy reduces once more to the form of Eq.~\eqref{eqn: def energy density}, such that we can still apply Eq.~\eqref{eqn: mean field energy}. On the other hand, the LHY energy can now be evaluated to yield \cite{petrov2015,ancilotto2018}
\begin{align}
\epsilon_{LHY}^{3D} = \frac{8}{15 \pi^2 \hbar^{3}} m_1^{3/2} (g_{11} n_1)^{5/2} f\left(\frac{m_2}{m_1},\frac{g^2_{12}}{g_{11}g_{22}},\frac{g_{22}n_2}{g_{11}n_1}\right),
\end{align}
with the function $f(1,x,y)$ defined as 
\begin{equation}\label{eqn: f(1xy)}
   f(1,x,y) = \sum_\pm \frac{1}{4\sqrt{2}}\qty(1+y \pm \sqrt{(1-y)^2+4xy})^{5/2}.
 \end{equation}
Close inspection of Eq.~\eqref{eqn: f(1xy)} reveals a problem known to exist upon deriving the LHY energy using Bogoliubov theory; for $\delta g <0$, the solution to $f(1,x,y)$, and therefore the LHY energy, becomes slightly complex \cite{petrov2015,petrov2023}. This is a consequence of the lack of a stable ground state solution in this regime at the MF level, which renders Bogoliubov theory strictly non-valid \cite{ota2020}.  However, because the dynamics of the instability are much slower than the dynamics of the stabilizing LHY term, the theory can still be applied \cite{petrov2023}. Since droplets exist in the regime where $\abs{\delta g} \ll g$, $\delta g$ can be set to zero, making the energy again a real number (which is also necessary not to exceed the accuracy of the Bogoliubov approximation \cite{petrov2023}). Then, considering equal mass components, the final expression for the LHY energy density in 3D corresponds to \cite{petrov2015}
\begin{align}\label{eqn: LHY energy final}
   \epsilon_\mathit{LHY}^{3D} = \frac{8 m^{3/2}}{15\pi^2 \hbar^3} (\gamma g_{22})^{5/2}n^{5/2}.
 \end{align}
It is important to mention that only for the LHY energy, $\delta g$ can be set to zero. $\delta g$ can not be set to zero in the MF energy (Eq. (\ref{eqn: mean field energy})), since it dominates the behavior of the term.\\

Compared to the analysis of the 3D energy density, the computation of the LHY energy for 1D systems is much more straightforward, as the momentum-space integration following from the continuum limit of Eq.~\eqref{eqn: LHY met g gewoon} is free of divergences in 1D. As such, the integral over $k$ can be directly computed, which results in the energy density
\begin{equation}
\label{eq:LHYdensity1D}
   \epsilon_{\mathrm{LHY}}^{1D}=\int\frac{dk}{4 \pi}\Bigg[E_{+,k}+E_{-,k}-\frac{\hbar^2 k^2}{2m_{\mathbf{r}}}-g_{11}n_1-g_{22}n_2 \Bigg]
   = - \frac{2\sqrt{m}}{3\pi \hbar} \qty(g_{11}n_1+g_{22}n_2)^{3/2},
 \end{equation}
 where it was assumed that $\delta g \ll g$ and that both components have the same mass ($m=m_1=m_2$). This result can also be rewritten as \cite{petrov2016}
 \begin{equation}\label{eqn: LHY 1D}
     \epsilon_{\mathrm{LHY}}^{1D} = - \frac{2\sqrt{m}}{3\pi\hbar}  \qty(\frac{g_{11}}{\gamma}n)^{3/2}.
 \end{equation}
The set of Eqs.~\eqref{eqn: mean field energy},~\eqref{eqn: LHY energy final} and~\eqref{eq:LHYdensity1D} provides us with all input required to obtain the eGPE, which we will proceed to analyze in the following subsection.

\subsection{Droplet ground state solutions to the eGPE}\label{sec eGPE}
To find the ground state solution for the 3D and 1D order parameters $\psi_j$, the stationary condition can be applied to the action functionals that govern these fields up to the level of the LHY corrections, this results in \cite{pitaevskij2016,petrov2015}
\begin{align} \label{eqn: egpe how to derive}
i\hbar\pdv{\psi_j(\vec{r},t)}{t} = \fdv{E}{\psi_j^*(\vec{r},t)}\Leftrightarrow  
   i\hbar\pdv{\psi_i}{t} 
   =- \frac{\hbar^2}{2m}\laplacian \psi_i  + \pdv{(\epsilon_\mathit{MF} + \epsilon_\mathit{LHY})}{n}\psi_i-\mu_i \psi_i,
\end{align}
where we have used $n = |\psi_1|^2+|\psi_2|^2$ and assumed that $m=m_1=m_2$. For 1D quantum droplets, the forms of the mean-field and LHY energy densities allow for an analytical solution of the order parameters. In contrast, such a solution is not accessible in 3D systems, where numerical methods must be employed. We therefore proceed by analyzing the 1D and 3D cases separately in the following subsections.
\subsubsection{Quantum droplets in 3D}
Substituting the 3D expressions for the mean-field and LHY energy densities as given by Eqs.~\eqref{eqn: mean field energy} and \eqref{eqn: LHY energy final} respectively, into Eq.~\eqref{eqn: egpe how to derive}, we obtain the following explicit form for the 3D eGPE \cite{petrov2015}
 \begin{equation}\label{eqn: gpe oude units}
   i\hbar\pdv{\psi_i}{t} =  \qty[- \frac{\hbar^2}{2m}\laplacian   - \frac{2\abs{\delta g} \gamma}{(1+\gamma)^2}n + \frac{4 m^{3/2}}{3\pi^2 \hbar^3} (\gamma g_{22})^{5/2}n^{3/2}-\mu_i]\psi_i.
 \end{equation}
In the Thomas-Fermi limit, the equilibrium density can be found by requiring that the total pressure is zero. Therefore, the following requirement is imposed
\cite{ferioli2015,ancilotto2018}
 \begin{equation}\label{eqn: P in 3D}
     P = - \pdv{\qty[(\epsilon_\mathit{MF}^{3D} + \epsilon_\mathit{LHY}^{3D}) V]}{V}\eval_{n_0} = 0,
 \end{equation}
 where, using Eq.~\eqref{eqn: def gamma}, the equilibrium density $n_0$ can be written in terms of the single component equilibrium density $n_{1,0}$ as 
$n_0 = \qty(1+\sqrt{\frac{g_{11}}{g_{22}}}) n_{1,0}$, from which it follows that \cite{petrov2015}
 \begin{equation}\label{eqn: infinite density}
     n_{i,0} = \frac{25 \pi^4 \hbar^6}{16m^3}\frac{(\delta g)^2}{g_{11}g_{22} \sqrt{g_{ii}}(\sqrt{g_{11}}+\sqrt{g_{22}})^5}
 \end{equation}
 This density can now be used to define the new length unit $\Tilde{\vec{r}} = \vec{r}/\xi$ and time unit $\tilde{t} = t/\tau$ where \cite{petrov2015}
 \begin{equation}\label{eqn units xi and tau}
   \xi=\sqrt{\frac{3\hbar^2}{2m}\frac{\sqrt{g_{22}}+\sqrt{g_{11}}}{|\delta g|\sqrt{g_{11}}n_{1,0}}} \qq{and} \tau=\frac{3\hbar}{2}\frac{\sqrt{g_{11}}+\sqrt{g_{22}}}{|\delta g|\sqrt{g_{11}}n_{1,0}}.
 \end{equation}
These new units and the rescaled order parameter $\psi = \psi_i/\sqrt{n_{i,0}}$ can now be introduced in the eGPE (Eq. (\ref{eqn: gpe oude units})) to find the dimensionless 3D eGPE \cite{petrov2015}
 \begin{equation}\label{eqn: egpe final}
   i\pdv{\psi}{\tilde{t}} = \qty(-\frac{1}{2}\laplacian_{\tilde{r}} - 3 \abs{\psi}^2 + \frac{5}{2}\abs{\psi}^3  -\tilde{\mu} )\psi.
 \end{equation}
Contrary to Eq.~\eqref{eqn: gpe oude units}, there are no longer two coupled equations that have to be solved to obtain the order parameters. It is useful to introduce a new rescaled number of atoms \cite{petrov2015}
\begin{equation}
\tilde{N} = \int \abs{\psi}^2  d^3\tilde{r} = \frac{N_i}{n_{i,0}\xi^3},
\end{equation}
 where $N_i$ is the number of atoms of one of the components. The energy units also change, resulting in the new energy
  \begin{equation}\label{eqn Etilde}
   \tilde{E} = \int \qty[\frac{1}{2}\qty(\nabla \psi)^2-\frac{3}{2}\abs{\psi}^4+\abs{\psi}^5] \dd\vec{\tilde{r}}.
   \end{equation}
A more detailed description of these transformations can be found in appendix \ref{appendix: units}.
The ground state can now be determined using an imaginary time evolution.
The convergence rate of this iterative procedure depends on the quality of the initial guess.
For smaller droplets, the Gaussian ansatz of the following form leads to fast converging times
\cite{hu2020}
 \begin{equation}
   \psi = \frac{\sqrt{N}}{\pi^{3/4}\sigma^{3/2}} \exp{-\frac{r^2}{2\sigma^2}},
 \end{equation}
where $\sigma$ is a variational parameter. The corresponding energy can be derived using Eq. (\ref{eqn Etilde}), which results in 
 \begin{equation}
   \frac{\tilde{E}}{\tilde{N}} = \frac{3}{4 \sigma^2} - \frac{3\tilde{N}}{4\sqrt{2}\pi^{3/2} \sigma^3} + \frac{4 \tilde{N}^{3/2}}{5\sqrt{10}\pi^{9/4}\sigma^{9/2}}.
 \end{equation}
By minimizing this energy the correct $\sigma$ can be found. For this model it turns out that below the critical particle number $\tilde{N} \approx 19.62$, there is no solution for $\sigma$, between $\tilde{N}\approx19.62$ and $\tilde{N}\approx24.03$ the solutions are metastable and above $\tilde{N}\approx 24.03$, the solution is stable \cite{hu2020}.  \\

For larger droplets, the Gaussian ansatz is outperformed by the numerically faster logistic ansatz, or Boltzmann function. This ansatz corresponds to \cite{alba-arroyo2022}
 \begin{equation}
   \psi = \frac{\psi_0}{1+\exp{\frac{r-R_0}{S}}},
 \end{equation}
where $\psi_0$ represents the maximum density, $S$ is related to the slope of the function and $R_0$ represents the $r$-coordinate where the function equals $\psi_0/2$. Here, $\psi_0$ and $S$ are carefully chosen by optimizing numerical calculations, whilst $R_0$ is determined through normalization of the wavefunction. \\

\begin{figure}[H]
    \centering
    \includegraphics[width=\textwidth]{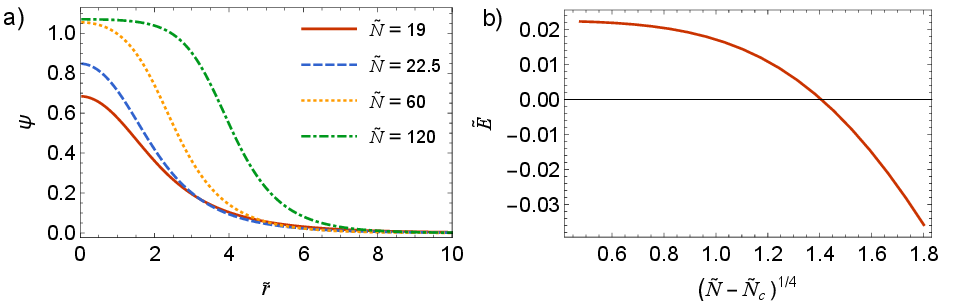}
    \caption{In Fig. a, the ground state order parameter for a three-dimensional quantum droplet as a function of the radial distance $\tilde{r}$ is plotted for different atom numbers $\tilde{N}$. In Fig. b, the ground state energy for a three-dimensional quantum droplet ($\tilde{E}$) as a  function of the number of atoms is plotted, where $\tilde{N}_c$ is the critical particle number.}
    \label{fig:groundstate}
\end{figure}

In Fig.~\ref{fig:groundstate}a the order parameter obtained from the imaginary time propagation is given. The obtained droplet stability of the numeric model is qualitatively similar to that of the Gaussian model. There is a critical particle number of $\tilde{N}_c\approx 18.65$, below which there are no stable droplet solutions \cite{petrov2015}. From the energy plot in Fig.~\ref{fig:groundstate}b it is clear that the droplets are only metastable for atom numbers between  $\tilde{N}_c$ and $\tilde{N}_m \approx 22.5$, before finding stable solutions for $\tilde{N} >\tilde{N}_m$ \cite{petrov2015}. It is also clear from Fig.~\ref{fig:groundstate}a that for small droplets, the solution is Gaussian-like, while the large droplets instead have a flat top profile.
\subsubsection{Quantum droplets in 1D}
To obtain the ground state solution for the 1D quantum droplets, the expressions for the mean-field and LHY energy densities as given by Eqs.~\eqref{eqn: mean field energy} and ~\eqref{eqn: LHY 1D} are substituted respectively into Eq.~\eqref{eqn: egpe how to derive}, finding 
\begin{equation}\label{eqn egpe 1D with units}
    i\hbar \pdv{\psi_i}{t} = \qty[-\frac{\hbar^2}{2m} \frac{\partial^2}{\partial r^2}  + \frac{2\delta g \gamma}{(1+\gamma)^2}n + \frac{1}{\pi\hbar}\sqrt{m}\qty(\frac{g_{11}}{\gamma})^{3/2}\sqrt{n}-\mu_i]\psi_i.
\end{equation}
Then, by once more applying the condition of zero pressure in the Thomas-Fermi limit, we find that the 1D equilibrium density corresponds to 
\cite{petrov2016}
\begin{equation}\label{eqn infinite density 1D}
    n_{0} = \qty[\frac{\sqrt{m}}{3\pi\hbar}\frac{g_{11}^{3/2}}{\delta g}\frac{(1+\gamma)^2}{\gamma^{5/2}}]^2 \Leftrightarrow n_{i,0}= \qty[\frac{\sqrt{m}}{3\pi\hbar}\frac{\sqrt{g_{ii}}g_{jj}}{\delta g}\frac{(1+\gamma)^{3/2}}{\gamma^{1/2}}]^2 \qq{with} j\neq i.
\end{equation}
Using the single component equilibrium density to rescale our wavefunction as $\psi = \psi_i/\sqrt{n_{i,0}}$, and introducing the dimensionless units $\tilde{r} = r/\xi$ and $\tilde{t} = t /\tau$, where \cite{astrakharchik2018a,debnath2023}
\begin{equation}\label{eqn: 1d untis}
    \xi =\frac{\pi\hbar^2\sqrt{\delta g}}{\sqrt{2}m g^{3/2}} \qq{and} \tau = \frac{\pi^2 \hbar^3\delta g}{2 m g^3}  
\end{equation}
the rescaled 1D eGPE can be rewritten as \cite{astrakharchik2018a,debnath2023} 
\begin{equation}\label{eqn: eGPE 1D dimensionless}
    i \frac{\partial \psi}{\partial \tilde{t}} = \qty[-\frac{1}{2} \frac{\partial^2}{\partial r^2} + \frac{4}{9}|\psi|^2  - \frac{2}{3}|\psi|-\tilde{\mu}] \psi,
\end{equation}
where equal intraspecies interaction strengths $g = g_{11} = g_{22}$ and equal masses $m=m_1=m_2$ are assumed. 
Just as in the 3D case, the rescaling with the equilibrium density has allowed us to transform two coupled eGPEs to a single equation. However, contrary to the 3D case, the 1D expression presented above can be solved analytically. Obtaining the time-independent version of the eGPE by substituting $\psi(r,t) =\psi(r)$, results in the following exact solution \cite{petrov2016,astrakharchik2018a,debnath2023}
\begin{equation}\label{eqn: 1d psi}
    \psi(\tilde{r},\tilde{\mu}) = -\frac{9\tilde{\mu}}{2\qty[1 + \sqrt{1 + \frac{9\tilde{\mu}}{2}} \cosh(\sqrt{-2\tilde{\mu}}\tilde{r})]},
\end{equation}
where chemical potential $\tilde{\mu}$, relates to the new rescaled number of atoms ($\tilde{N}$) via the relation \cite{petrov2016,astrakharchik2018a,debnath2023}
\begin{equation}\label{eqn: mu to N}
    \tilde{N} = \int_{-\infty}^{+\infty} |\psi(\tilde{r})|^2 \, d\tilde{r} = 3 \left[ \ln\left( \frac{3\sqrt{-\tilde{\mu}/2 }+ 1}{\sqrt{1 + \frac{9\tilde{\mu}}{2}}} \right) - 3\sqrt{-\tilde{\mu}/2} \right].
\end{equation}
From the previous expression it follows that, if $\tilde{\mu}=0$, then there are no particles ($\tilde{N}=0$) and, if $\tilde{\mu}=-2/9$, then there are an infinite number of particles ($\tilde{N}\rightarrow  \infty$). Given the number of atoms $\tilde{N}$, Eq. (\ref{eqn: mu to N}) can be solved numerically to obtain $\tilde{\mu}$, which then can be used to find the order parameter from Eq. (\ref{eqn: 1d psi}).
Just as for the 3D droplets was the case, the units for the energy will again change, resulting in the new expression for the energy
\begin{equation}\label{eqn Etilde 1D}
    \tilde{E} = \int_{-\infty}^{+\infty} \left( \frac{1}{2}  \left| \frac{\partial \psi}{\partial \tilde{r}} \right|^2 + \frac{2}{9} |\psi|^4 - \frac{4}{9}  |\psi|^3 \right) d\tilde{r}
\end{equation}
 However since a droplet has to exist out of at least two particles, the energy will be set to zero if there is only one atom in the droplet.\\

\begin{figure}[H]
    \centering
    \includegraphics[width=\textwidth]{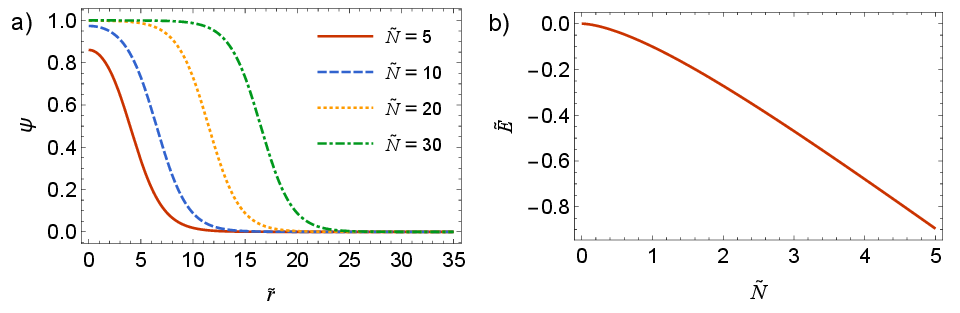}
    \caption{Fig. a shows the ground state order parameter for a one-dimensional quantum droplet as a function of the distance $\tilde{r}$ for different atom numbers $\tilde{N}$. Fig. b shows the ground state energy for a one-dimensional quantum droplet ($\tilde{E}$) as a function of the number of atoms. }
    \label{fig:groundstate1D}
\end{figure}

In Fig.~\ref{fig:groundstate1D}a the order parameter as a function of the distance $\tilde{r}$ is given for various rescaled particle numbers.  Similar to a 3D droplet, small 1D droplets have a Gaussian-like shape and bigger droplets have a flat top shape. An important difference compared with 3D droplets is that there is no minimal number of atoms required to form stable 1D droplets. As is clear from Fig.~\ref{fig:groundstate1D}b, the 1D ground-state energy is always negative, and the droplet is thus stable at zero temperature for arbitrary particle number.

\section{Theory of droplet fission}\label{sec:Theory of droplet fission}
Fragmentation of a droplet into multiple smaller droplets leads to an increase in the system’s energy, but also results in a gain in entropy. As temperature rises, the entropy contribution becomes increasingly significant. At a certain critical temperature, referred to as the fragmentation temperature ($T_f$), the entropy gain can outweigh the energy cost, making droplet fragmentation thermodynamically favorable. A similar phenomenon has been explored in the context of solitons in a single-component one-dimensional Bose gas \cite{herzog2014}. In this work, we adopt a comparable framework to investigate the fragmentation behavior of 3D as well as 1D quantum droplets. Focusing on dilute systems where the two components are in balance and have equal intraspecies interaction strengths, it can be assumed that the droplets are non-interacting. This is because the droplets can only interact via local-interactions. Another way the droplets could interact, if they were allowed to be excited, is via the shape oscillations and the breathing mode of the droplets. The shape oscillations would lead to interactions ranges up to the diameter of the droplet. The breathing mode could cause interaction with longer interaction ranges. However, in order to do so, there has to be a medium between the droplets, which we do not have. Even if there was some residual gas, because the components are not balanced, the medium would be compressible and lessen the effect compared to bubbles in a liquid. It is thus safe to assume that the ground state can be described by the partition function $Z_{ideal}$ of an ideal gas, where \cite{sethna2021}
\begin{align}
\label{eq:Zideal}
Z_{ideal} = \frac{1}{N!}\left(\frac{V}{\lambda^D}\right)^N, \qquad \textrm{where} \qquad \lambda = \frac{h}{\sqrt{2\pi m  k_B T}}.
\end{align}
Here, $D$ represents the dimension and $N$ represents the total particle number.
However, contrary to an ideal gas, the constituents of our system are droplets instead of atoms. Considering droplets consisting of $n$ atoms (with $N_c \leq n \leq N$), the total number of atoms $N$ in these systems is fixed by $\sum^N_{n = N_c} n \mathcal{N}_n = N$, where $\mathcal{N}_n$ represents the total number of droplets with $n$ atoms\footnote{From section \ref{sec:Theory of droplet fission} onwards $n$ will no longer denote a total density.}. Then, the Broglie wavelength associated with these droplets can be defined as $\lambda_{droplet} =  \lambda/\sqrt{n}$. The partition function of a gas that contains $\mathcal{N}_n$ droplets with $n$ atoms, corresponds to the external partition function

\begin{equation}\label{eqn: external partiotn function}
  Z_{\mathcal{N}_n}^e = \frac{1}{\mathcal{N}_n!}\qty(\frac{V}{\lambda^D} \qty(\frac{N}{\mathcal{N}_n})^{D/2})^{\mathcal{N}_n}.
\end{equation}
The partition function of a configuration with different sized droplets, corresponds to the multiplication of all $Z_{\mathcal{N}_n}^e$ in the configuration, weighted by the internal partition function $Z^i_{\mathcal{N}_n}$, which is the Boltzmann factor
\begin{align}
Z^i_{\mathcal{N}_n} = \exp\{-\beta \mathcal{N}_n E(n)\}.
\end{align}
The total partition function is the sum over all possible configuration, and thus equals
\begin{equation}\label{eq: Z meest algemeen}
    Z = \sum_{\mathrm{config}} Z_{\mathrm{conf}} = \sum_{\mathrm{config}}\left[\prod_{n=N_c}^N Z^e_{\mathcal{N}_n}Z^i_{\mathcal{N}_n}\right].
\end{equation}
Where the energy $E(n)$ corresponds to the single droplet energy that was derived in section \ref{sec eGPE}.
From the partition function, we focus on extracting and investigating two key properties. First of all, we compute the free energy ($F = -k_B T \ln[Z]$) to determine the thermodynamic ground state of the system. Next, the average fragmentation rate $\langle\nu\rangle$ is defined as 
\begin{align}
\langle\nu\rangle = \frac{1}{Z}\sum_{\mathrm{config}} \nu(\{\mathcal{N}_n\}) Z_{\mathrm{conf}}
\end{align}
with
\begin{equation}\label{eqn: v met floor}
    \nu(\{\mathcal{N}_n\}) = \sum_{n = N_c}^N \frac{\mathcal{N}_n}{\mathcal{N}_n^\mathit{max}} \qq{and} \mathcal{N}_n^\mathit{max} = \text{Floor}\qty(\frac{N}{N_c})
\end{equation}
From the above definition, it follows that the average fragmentation rate is the statistically averaged value of $\nu(\{\mathcal{N}_n\})$, which is equal to $1/\mathcal{N}_n^\mathit{max}$ if there is a single droplet that contains all $N$ atoms, and equals one if the droplet is fully fragmented into droplets of (the minimum) size $N_c$. As such, the fragmentation rate $\langle\nu\rangle$ tells us, if a droplet is fragmented, how many fragments exist on average.\\

It should be noted that computing the full partition function $Z$ is generally a hard problem. Particularly for 3D systems and systems with large particle number, the number of possible configurations will be large and numerical calculations will be computationally demanding. To limit the complexity, we obtain an estimate for the transition temperature and fragmentation rate by computing the partition function only for configurations where all (fragmented) droplets are of equal size.
Then, by adding more configurations (such as one where two droplets have merged), we can check how much our estimate is affected and thereby evaluate the validity of our approximation.  Under the approximation of equal sized droplets the partition function as defined in Eq. (\ref{eq: Z meest algemeen}) simplifies to 
\begin{equation}\label{eqn: Z met eenheden opgelost}
  Z = \sum_{\mathcal{N}_n} e^{-\tilde{\beta} \mathcal{N}_n \tilde{E}(n)} \frac{1}{\mathcal{N}_n!}\qty[\frac{\tilde{V}}{\tilde{\lambda}^D} \qty(\frac{2 n_{i,0}\xi^3\tilde{N}}{\mathcal{N}_n})^{D/2}]^{\mathcal{N}_n} 
  \qq{where} \frac{\tilde{N}}{\mathcal{N}_n}>\tilde{N}_c.
\end{equation}
Furthermore, the average fragmentation rate will also simplify to
\begin{equation}
\label{eqn:unitsnuav}
    \ev{\nu} = \frac{1}{Z}\sum_{\mathcal{N}_n=1}^{\mathcal{N}_n^\mathit{max}} \frac{\mathcal{N}_n}{\mathcal{N}_n^\mathit{max}}  Z_{\mathcal{N}_n} \qq{where} \mathcal{N}_n^\mathit{max} = \text{Floor}\qty(\frac{N}{N_c}).
\end{equation}
In Eqs.~\eqref{eqn: Z met eenheden opgelost} and~\eqref{eqn:unitsnuav} the units of equilibrium density $n_{i,0}$ are used as previously introduced in Sec.~\ref{sec eGPE} and elaborated on in App.~\ref{appendix: units}. This makes it possible to directly use the energies derived in Eqs. (\ref{eqn Etilde}) and (\ref{eqn Etilde 1D}). Following the strategy applied in Ref. \cite{herzog2014} to study solitons in 1D Bose gases, we obtain further analytical estimates for the fragmentation temperature and fragmentation rate by considering only the limiting case where either all atoms are within one droplet, with free energy $\tilde{F}_1$ or where the droplet is maximally fragmented, with free energy $\tilde{F}_{\mathcal{N}_n^\mathit{max}}$. 
These energies can be computed by considering a single term in Eq.~\eqref{eqn: Z met eenheden opgelost}, such that 
\begin{multline}
    \tilde{F}_{\mathcal{N}_n} = -\frac{1}{\tilde{\beta}}\ln Z_{\mathcal{N}_n} \\
    = -\frac{1}{\tilde{\beta}}\left[ - \tilde{\beta} \mathcal{N}_n \tilde{E}(n) - \ln \{ \mathcal{N}_n ! \} + \mathcal{N}_n \ln \left\{ \frac{\tilde{V}}{\tilde{\lambda}^D} \right\} + \frac{D}{2} \mathcal{N}_n \ln  \left\{  \frac{2 n_{i,0}\xi^D\tilde{N}}{\mathcal{N}_n}  \right\}  \right],
\end{multline}
By comparing $\tilde{F}_1$ and $\tilde{F}_{\mathcal{N}_n^\mathit{max}}$ and finding the point where the free-energy of the fully fragmented state becomes smaller than the free energy of the single droplet state, an estimate can be found for the fragmentation temperature ($T_f$)
\begin{equation}\label{eqn: 3D frag temp}
    \tilde{\beta}_f = \frac{
    -\frac{D}{2} \left( 
        \mathcal{N}_n \ln \left\{ \frac{2 n_{i,0}\xi^D\tilde{N}}{\mathcal{N}_n} \right\} 
        - \ln \{2 n_{i,0}\xi^D\tilde{N}\} \right)
        + \ln\{\mathcal{N}_n!\} 
        + (1 - \mathcal{N}_n) \ln\left\{\frac{\tilde{V}}{\tilde{\lambda}_f^D}\right\} 
}{
    \tilde{E}(\tilde{N}) - \mathcal{N}_n \tilde{E}(\tilde{n})
},
\end{equation}
where $\tilde{\lambda}_f$ is, according to Eq. (\ref{eqn: broglie wavelength tilde}), explicitly temperature dependent. For 1D droplets the above relation simplifies further since the critical particle number $N_c$ equals one, such that
\begin{equation}\label{tilde beta c 1d}
        \tilde{\beta}_f^{1D} = \frac{
    \frac{1}{2} 
      \ln \{2 n_{i,0}\xi^D\tilde{N}\} 
        + \ln\{\mathcal{N}_n!\} 
        + (1 - \mathcal{N}_n) \ln\left\{\frac{\tilde{L}}{\tilde{\lambda}_f}\right\} 
}{
    \tilde{E}(\tilde{N})
}.
\end{equation}
In the next section we will proceed to compute the fragmentation temperature and average fragmentation rate for both 3D and 1D systems for various interaction strengths and carefully analyze the various approximations to the partition function  introduced in this section.\\

Since we have access to the full configuration for 1D droplets, it is possible to accurately calculate the specific heat,
\begin{equation}\label{formule: Cv}
    C_V = -T \pdv[2]{F(T,L,N)}{T}\eval_{L,N}.
\end{equation}
This will be a very useful quantity since it will be sharply peaked at the transition, and can be used to define the fragmentation temperature.

\section{Results and discussion}\label{sec: results}
Both in the three- and one-dimensional case, the fragmentation temperature will be compared with the critical temperature ($T_C$) \cite{guebli2021,boudjemaa2023a}. Above this temperature, quantum droplets can not exist in an infinite system, and the system will transition back into an expanding mixture. Our current model does not describe this state. The critical temperature is calculated by calculating the free energy of a single droplet in an infinite volume as a function of the density. $T_C$ corresponds with the temperature where the local minima that corresponds with the droplet solution (density) disappears. Even though our system is not infinite, $T_C$ will be used to define where these droplets can or cannot exist. In a finite system we expect $T_C$ to be higher.\\

In this section different units for 3D droplets and 1D droplets will be introduced. These units are therefore introduced in each subsection. 

\subsection{Results and discussion for 3D droplets}
For 3D droplets, the energy will be expressed in units of 
\begin{equation}
    E_0 = \frac{\hbar^2}{ma^2},
\end{equation}
Where $a=a_{11}=a_{22}$. This results in the fact that the temperature and free energy will scale as
\begin{equation}
    E_0 \beta \sim \sqrt{\frac{g}{\abs{\delta g}}} \qq{and} \frac{F}{E_0} \sim  \sqrt{\frac{\abs{\delta g}}{g}}.
\end{equation}
This is worked out in more detail in Appendix \ref{appendix: units}.\\

For three-dimensional quantum droplets, there are three approximations for the configurations used. One is where each configuration exists out of the same sized quantum droplets, which corresponds with the partition sum in Eq. (\ref{eqn: Z met eenheden opgelost}). In the second one, only two configurations are taken into account, one where all atoms are within one droplet and one where the droplet is maximally fragmented. In the third, again the two extreme configuration are taken into account, but also the states surrounding them. For the state that contains one droplet, two states are added, one with two droplets containing $N-N_c$ and $N_c$ atoms, and one with two droplets containing $N-N_m$ and $N_m$ atoms.
For the fully fragmented state, the state where two droplets have merged is added. It turns out that these three models, differ by a negligible amount.\\

\begin{figure}[!htb]
    \centering
    \includegraphics[width=0.6\linewidth]{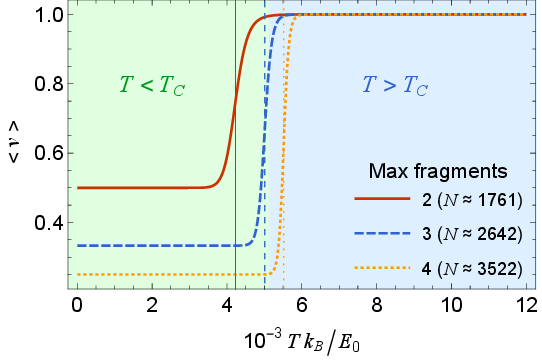}
    \caption{The average fragmentation rate ($\ev{v}$) of 3D droplets for the equal part configuration that is derived in Eq. (\ref{eqn:unitsnuav}), is given as a function of the temperature for different number of particles for $\abs{\delta g}/g = 0.4$ and $V = 10^{12} a^3$. The vertical lines corresponds to the fragmentation temperature, calculated with Eq. (\ref{eqn: 3D frag temp}). $T_C$ is the critical temperature \cite{guebli2021} and divides the green zone where droplets can exist, with the blue zone where they can not exist in the limit of $V\rightarrow\infty$. }
    \label{fig:3Dde3Modellen}
\end{figure}

In Fig. \ref{fig:3Dde3Modellen} the average fragmentation rate is given for the equal part configuration, that is derived in Eq. (\ref{eqn:unitsnuav}). The vertical lines correspond with the simplest approximation for the fragmentation temperature, defined in Eq. (\ref{eqn: 3D frag temp}). The models give such a similar result because the droplets have a minimum number of atoms ($N_c$). This means that splitting a droplet with $N$ atoms, in a slightly smaller droplet with $N-N_c$ atoms and a small droplet with $N_c$ atoms, requires a considerable amount of energy. Similarly, joining two droplets of $N_c$ atoms, changes the number of particles and energy too drastically.\\

\begin{figure}[!htb]
    \centering
    \includegraphics[width=\linewidth]{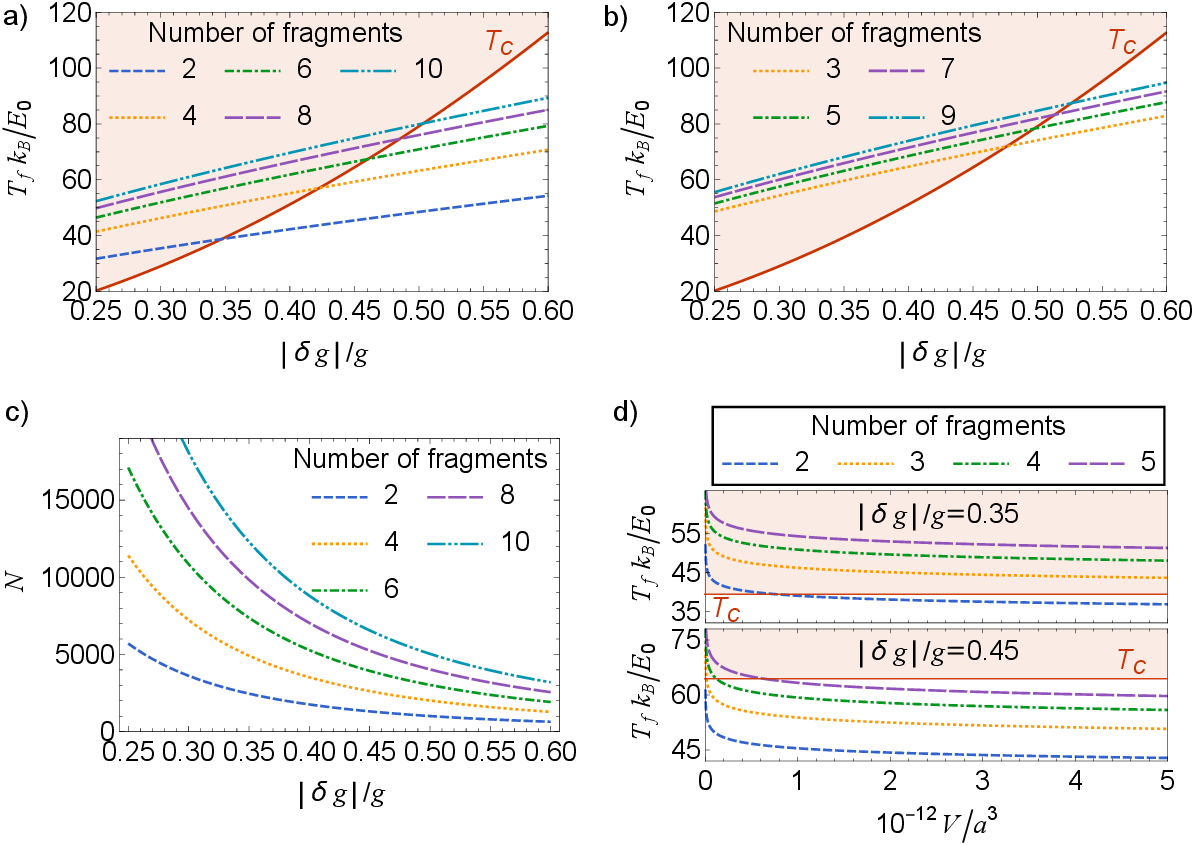}
    \caption{In Fig. a, the fragmentation temperature ($T_f$) is plotted for 3D droplets that are just able to split into smaller droplets that are slightly larger than $N_c$ ($\tilde{N}/\mathcal{N}_n^\mathit{max}=18.7\approx 18.65 = \tilde{N}_c$). In Fig. b, the fragmentation temperature ($T_f$) is plotted for droplets that are just not able to split into smaller droplets with $N_c$ atoms ($\tilde{N}/(\mathcal{N}_n^\mathit{max}+1) = 18.6 \approx \tilde{N}_c$). In both figures, $T_f$ is given for different droplet sizes. These droplets will thus split in different numbers of fragments ($\mathcal{N}_n^\mathit{max}$). In both figures $T_f$ is plotted as a function of $\abs{\delta g}/g$ for a system with a volume $V=10^{12}a^3$. In Fig. c, the corresponding number of atoms is given for Fig. a. There is no visual difference with the number of atoms for Fig. b, if you compare the same line types (e.g. compare 3 in Fig. b with 4 in Fig. c). In Fig. d, $T_f$ is given as a function of the system size ($V/a^3$) for droplets that are just able to split into smaller droplets and for two different values of $\abs{\delta g}/g$. The red lines in these plots correspond with the critical temperature \cite{guebli2021}. In the red zone where $T>T_C$, no droplets can exist in an infinite system.}
    \label{fig:resultaten3D}
\end{figure}

In Fig. \ref{fig:resultaten3D}(a-b), the fragmentation temperature is plotted as a function of $\abs{\delta g}/g$ for droplets that can just (a), just cannot (b), fragment into different sizes. If the number of atoms increases, the entropy will increase which would lower $T_f$. However, the energy difference between the fragmented state and one-droplet state will also increase, which increases $T_f$. It turns out that the latter is a stronger effect, and that $T_f$ will increase if the amount of atoms increases. This means that $T_f$ will be the lowest if the droplet has $2 N_c$ atoms, so that the droplet has just enough atoms to split into two. Increasing $N$ also results in jumps for the fragmentation temperature. These occur because of the floor function in Eq. (\ref{eqn: v met floor}). If $N$ increases there will be a discontinuous jump in this floor function, which then results in a jump in the fragmentation temperature. This is also clear from comparing Figs. \ref{fig:resultaten3D}a and \ref{fig:resultaten3D}b.  Since increasing $\abs{\delta g}/g$ results in an increase in the energy (see Eq. (\ref{eqn: 3D voorfactor})), the fragmentation temperature will also increase. Increasing $\abs{\delta g}/g$ also increases the critical temperature. However, the critical temperature increases faster. This results in droplets that do not fragment for small $\abs{\delta g}/g$, but do fragment for higher $\abs{\delta g}/g$. Increasing $V$ will decrease the fragmentation temperature since it will increase the entropy. This decrease is initially very fast, but slows down drastically if the system has reached a considerable size. This means that for realistic systems, droplets will not fragment or will only fragment for small droplet sizes if $\abs{\delta g}/g$ is too small.\\

Our energy functional yields lower energies than those obtained using HFB calculations \cite{guebli2021} or DMC simulations \cite{cikojevic2019a}, this means that fragmentation will occur at lower temperatures than our theory predicts, and we therefore provide a conservative upper bound for the fragmentation temperature.

\subsection{1D droplets}\label{sec: 1Dresults}
For 1D droplets, the energy will be expressed in units of the binding energy of an interspecies dimer
\begin{equation}
    E_B = -\frac{\hbar^2}{ma_{12}^2}.
\end{equation}
This results in the fact that the temperature and free energy will scale as
\begin{equation}
    \abs{E_B} \beta \sim \qty(\frac{\delta g}{g})^{5/2}\qty(1-\frac{\delta g}{g})^{2} \qq{and} \frac{F}{\abs{E_B}} \sim  \qty(\frac{g}{\delta g})^{5/2}\qty(1-\frac{\delta g}{g})^{-2}.
\end{equation}
For 1D droplets it is also important to mention that $\xi \sim \sqrt{\frac{\delta g}{g}}\abs{a}$. This means that if $\delta g/g$ increases, also the size of the droplet will increase. This is worked out in more detail in Appendix \ref{appendix: units}.\\

\begin{figure}[!htb]
    \centering
    \includegraphics[width=0.8\linewidth]{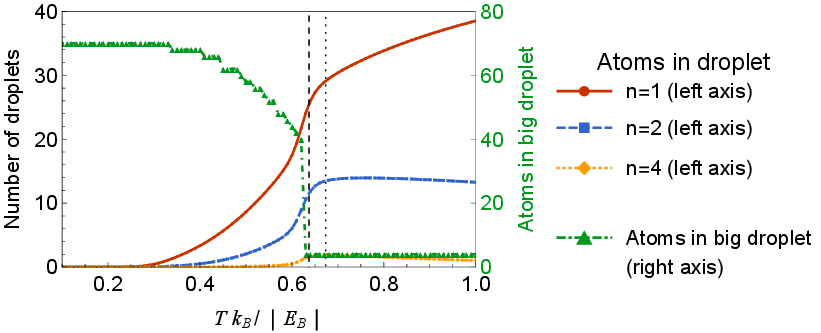}
    \caption{In this figure the ground state of a 1D droplet is shown as a function of the temperature ($T$) for a system with 70 atoms a size $L=1000 \abs{a}$ and $\delta g/g=0.15$. The green line corresponds with the number of atoms in the biggest droplet of the most contributing configuration (the configuration with the highest $Z_\text{conf}$). The other lines correspond with the expectation value of the amount of droplets that have one, two or four atoms. The vertical dotted line corresponds with the approximation for the fragmentation temperature derived in Eq. (\ref{tilde beta c 1d}). The vertical dashed line corresponds with the more complex approximation described in section \ref{sec: 1Dresults}.}
    \label{fig:3moddellen}
\end{figure}

As opposed to 3D quantum droplets, droplets in 1D can form for arbitrary small atom number. Therefore, contrary to 3D droplets, it is numerically feasible to simulate the partition function for the full configuration for small 1D droplets. The full configuration can be generated by modifying an algorithm of \cite{stockmal1962}. By studying the most occupied states for different temperatures, which corresponds to the states with the highest $Z_\mathit{conf}$ in Eq. (\ref{eq: Z meest algemeen}), the transition can be understood. In Fig. \ref{fig:3moddellen} we analyze the ground state of 1D droplets as a function of temperature. Here we recognize that an increase of temperature results in atoms being expelled from the droplet. These expelled atoms mostly form a gas, or pairs (droplets of 2 atoms), but also droplets with 4, 6 or more atoms. A similar behavior was observed for solitons \cite{herzog2014}, where also pairs stayed stable for higher temperatures.\\

Using the full configuration is very computationally demanding for bigger droplets. So instead, to study larger droplets, two approximations will be used. In the first approximation only the fully fragmented, or gas state, and the state with all atoms inside a single droplet is taken into account. Under this assumption, the fragmentation temperature $T_f$ is computed according to Eq. (\ref{tilde beta c 1d}). In the second approximation, the configuration contains one big droplet that can contain between 1 and $N$ atoms. All remaining atoms, form a gas that can have an arbitrary number of pairs. The first approximation works well for bigger systems and lower values of $\delta g/g$, the second approximation also works for smaller systems and higher $\delta g/g$, until also droplets with 4 and 6 atoms start to contribute to the fragmentation temperature. The transition temperature for the second approximation is found by calculating and determining the peak in the specific heat using Eq. (\ref{formule: Cv}).\\

\begin{figure}[!htb]
    \centering
    \includegraphics[width=\linewidth]{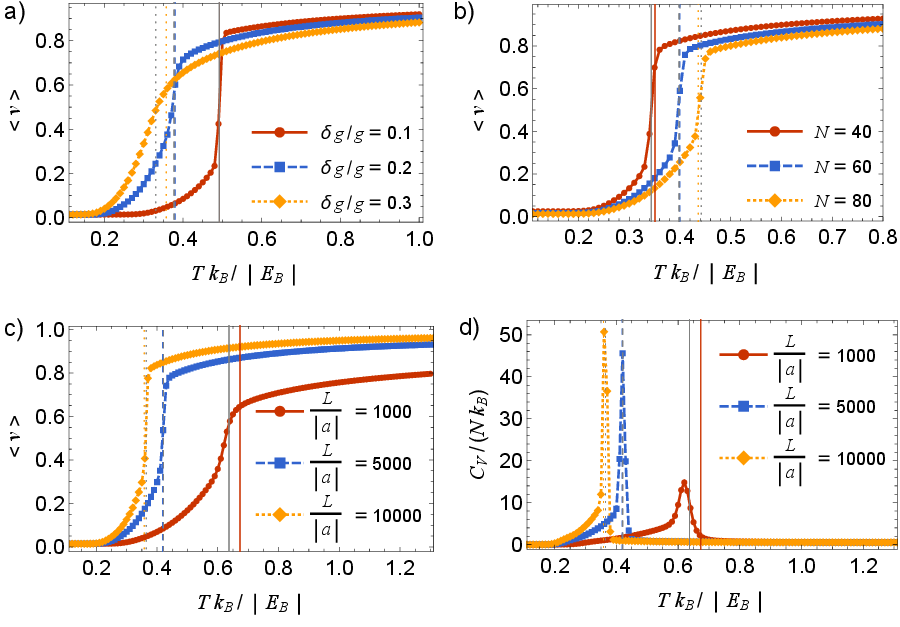}
    \caption{In Fig. a-c, the average fragmentation $\ev{v}$ is given as a function of the temperature $T$ for a 1D system. In Fig. a, $\ev{v}$ is plotted for different $\delta g/g$ where $L=5000 \abs{a}$ and $N=70$. In Fig. b this is plotted for different number of particles ($N$) where $L=5000 \abs{a}$ and $\delta g/g = 0.15$. In Fig. $c$ it is plotted for different system sizes ($L$) where $N=70$ and $\delta g/g=0.15$. In Fig. d, the specific heat ($C_V$) is given as a function of the temperature for different system sizes, where  $N=70$ and $\delta g/g=0.15$. In all the subfigures the curves correspond to the calculations with the full configurations. The vertical colored lines correspond with the approximation in Eq. (\ref{tilde beta c 1d}). The gray lines that are right next to the colored lines, correspond with the more complex approximation described in section \ref{sec: 1Dresults}.}
    \label{fig:1DResults}
\end{figure}

In Fig. \ref{fig:1DResults} we proceed with the analysis of the average fragmentation $\ev{\nu}$ as a function of temperature for different interaction strengths (Fig. \ref{fig:1DResults}a), number of atoms (Fig. \ref{fig:1DResults}b) and system sizes (Fig. \ref{fig:1DResults}c). Studying Fig. \ref{fig:1DResults}a, it is clear that if $\delta g/g$ increases, the fragmentation temperature will decrease. This can be understood from Eq. (\ref{eqn units 1D}), where we recognize that the energy decreases as a function of $\delta g/ g$. Similarly to the fragmentation temperature, the critical temperature $T_C$ is also sensitive to the ratio $\delta g/ g$ and will also decrease for increasing $ \delta g / g$. However, this decrease in $T_C$ is faster than the decrease in $T_f$, such that at some point the fragmentation temperature becomes larger than the critical temperature and the fragmentation of 1D droplets will no longer be observable. This is analyzed in more detail further on in this section. In line with Eq. (\ref{eq lengte 1D}), the average fragmentation becomes smoother for increasing $\delta g / g$. This is a result of the increase of the droplet size.\\

Studying Fig. \ref{fig:1DResults}b, it is clear that increasing the atom number, results in a smoother transition of the fragmentation rate, since an increase in atom number allows for more available states to be occupied. Similar to 3D droplets, an increase in the number of atoms also result in an increase of the fragmentation temperature, which can be similarly explained by the increase of energy being larger than the increase of the entropy for the addition of atoms to the system. On the other hand, increasing the system size as presented in Fig. \ref{fig:1DResults}c, results in a sharper transition at earlier temperatures. This behavior was also observed for 3D droplets and is explained by the increase of the entropy for larger systems. In Fig. \ref{fig:1DResults}d, the specific heat is also given. As can be expected for an ideal gas, it goes to $\frac{1}{2}k_B$ if there is one droplet and will converge to a value of  $\frac{1}{2} k_B N$ for the fully fragmented state.\\

\begin{figure}[!htb]
    \centering
    \includegraphics[width=\textwidth]{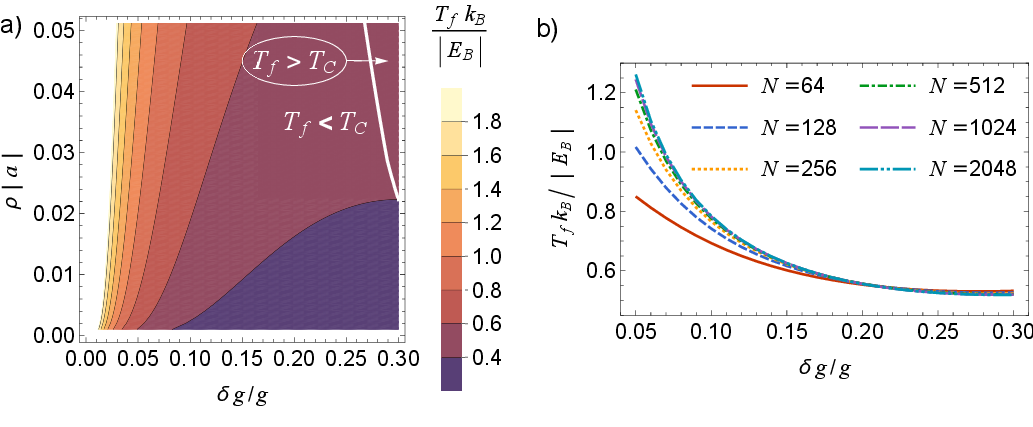}
    \caption{In Fig. a, the fragmentation temperature ($T_f$) in the thermodynamic limit (where $N$ is fixed to 2048 atoms) for a 1D droplet as a function of $\delta g/g$ and density ($\rho = N/L$) in units of the scattering length. The white line corresponds to the critical temperature of the droplets \cite{boudjemaa2023a}. In Fig. b, $T_f$ is given as a function of $\delta g/g$, for different system sizes with the same density of atoms ($\rho=0.0512\; \abs{a}$). $T_f$ is in both figures calculated using the approximation in Eq. (\ref{tilde beta c 1d}).}
    \label{fig:thermallimit1D}
\end{figure}

In Fig. \ref{fig:thermallimit1D}a, the fragmentation temperature is shown as a function of $\delta g/g$ and the density, where the system has reached the thermodynamic limit as evidenced by Fig. \ref{fig:thermallimit1D}b, where the fragmentation temperature as a function of $\delta g/ g$ saturates for sufficiently large system sizes. It is clear that the droplet will fragment in the thermodynamic limit as long as the density is not very high in combination with a high $\delta g/g$. We note that this figure is made with the most simple approximation for $T_f$ and is only qualitatively correct. Particularly, we expect the approximation to slightly underestimate the value of $T_f$.  \\ 

There are some important things to keep in mind. The approximation for the energy is in absolute value higher than in reality \cite{parisi2019}. A lower energy in absolute value will further lower the transition temperature. This analysis only studies the most stable ground state. It does not predict that this transition will spontaneously take place. $\delta g/g$ is also limited to a value of 0.3, because the model also qualitatively starts to fail compared to the Monte Carlo results \cite{parisi2019}.

\section{Conclusion and Outlook}\label{sec concl outlook}
In this work, we have shown that at non-zero temperatures the thermodynamic ground state for quantum droplets in 1D and 3D systems is not necessarily a single droplet, but can instead consist out of multiple droplets or a gas. As the temperature increases, not only the energy of the system will be important but also the entropy. This means that at a certain temperature it will be beneficial to increase the entropy by fragmenting the droplet in smaller droplets, even tough the ground state energy will be higher. However, this temperature may lie outside of the regime that these droplets can exist. In this work we verified that there is a big regime where the fragmentation temperature is within the stable regime of these droplets.\\

For three-dimensional droplets, this transition occurs when the system is sufficiently large, the ratio $\abs{\delta g}/g$ is high enough and has a low enough number of atoms. The lowest fragmentation temperature occurs when the droplet is just capable to split in two smaller droplets.\\

For one-dimensional droplets, the increase in temperature causes the droplet to expel atoms, which mainly form a gas of free atoms and droplets with two atoms, but larger droplets may also appear. These two atom droplets remain present even at relatively high temperatures compared with the fragmentation temperature itself. The fragmentation temperature will be lowered if $\delta g/g$ increases, the number of atoms decreases or the system size increases. For large values of $\delta g/g$, fragmentation may not occur at all.\\

This work can be further improved by introducing interactions between the droplets. However, these interactions are expected to have a small effect due to the low density and local nature of interaction. The model for the one-dimensional droplets could also be further refined by using a more accurate model for the energy \cite{parisi2019}. For three- and one-dimensional droplets the internal free energy of the droplets can also be taken into account by using HFB simulations \cite{guebli2021,boudjemaa2023a}. It would be interesting to study this fragmentation process dynamically to identify possible energy barriers \cite{tempere2003}. It would also be interesting to study the fragmentation in harmonic traps, especially since fragmentation was already observed in dynamical simulations of the system \cite{pathak2022,bristy2025}.\\

\backmatter

\bmhead{Acknowledgements}
J.V.L. gratefully acknowledges funding by the Research Foundation - Flanders (FWO-Vlaanderen), through doctoral (PhD) grant for fundamental research, grant number 1184125N. D.A.-B. acknowledges funding from the
Research Foundation-Flanders via a postdoctoral fellowship (Grant No. 1222425N). We acknowledge financial support by the Research Foundation Flanders (FWO), Projects No. G0AIY25N, No. G0A9F25N, and No. GOH1122N.

\begin{appendices}

\section{Unit transformations}\label{appendix: units}
In this paper, a lot of different units are used. In this appendix the transformation between these different units will be worked out in more detail. In section \ref{sec eGPE}, the new units
\begin{equation}
    \vec{\tilde{r}} = \frac{\vec{r}}{\xi} \qq{and} \tilde{t} = \frac{t}{\tau}
\end{equation}
are introduced. $\xi$ and $\tau$ have different expressions depending on the dimensionality of the system (D). For 3D droplets the definitions for $\xi$ and $\tau$ are given in Eq. (\ref{eqn units xi and tau}) and for 1D droplets in Eq. (\ref{eqn: 1d untis}). The order parameters are also rescaled by the equilibrium density $\psi=\psi_i/\sqrt{n_{i,0}}$, where $n_{i,0}$ is given in Eq. (\ref{eqn: infinite density}) for 3D droplets and in Eq. (\ref{eqn infinite density 1D}) for 1D droplets. This also results in a rescaled number of particles and energy. The number of particles follows from
 \begin{equation}\label{eqn tildeN appendix}
     \tilde{N} = \int \abs{\psi}^2  d^D\tilde{r} = \frac{N_i}{n_{i,0}\xi^D}.
 \end{equation}
The rescaled energy can be found by explicitly filling in the expression for the energy or by doing the unit transfer in Eq. (\ref{eqn: egpe final}) and (\ref{eqn: eGPE 1D dimensionless}) explicitly:
\begin{equation}
  i\hbar\pdv*{\psi_i}{t} =  \qty[\pdv{\epsilon}{n}]\psi_i \Leftrightarrow 
  i\pdv{}{\tilde{t}}\qty(\sqrt{n_{1,0}}\psi) = \frac{\tau}{\hbar}\Big[\pdv{\epsilon}{n}\Big]\qty(\sqrt{n_{1,0}}\psi)
  \Rightarrow
  \frac{\tau}{\hbar}\pdv{\epsilon}{n} = \pdv{\tilde{\epsilon}}{\tilde{n}},
\end{equation}
where the quantities with a tilde are the dimensionless/rescaled quantities and $\epsilon$ the energy density (including the kinetic term). From here, it will be assumed that $g_{11}=g_{22}=g$ and $m=m_1=m_2$, which also means that $n_{1,0}=n_{2,0}=n_{i,0}$ (see Eq. (\ref{eqn: def gamma})). Using Eq. (\ref{eqn tildeN appendix}) the density can be rewritten as $n=2 n_{i,0} \tilde{n}$. The energy density can then be written as $\epsilon = 2 n_{i,0} \frac{\hbar}{\tau} \tilde{\epsilon}$, so that finally the energy transforms as
\begin{equation}\label{eqn: tilde E ifv E}
    \tilde{E} = \frac{\tau}{\hbar} \frac{1}{2n_{i,0} \xi^D} E.
\end{equation}
For consistency the inverse temperature and free energy ($\tilde{F} = - \frac{1}{\tilde{\beta}}\ln \qty[Z]$) will be expressed in the same units
\begin{equation}
  \tilde{\beta} = \frac{\hbar}{\tau} (2n_{i,0} \xi^D) \beta \qq{and} \tilde{F} = \frac{\tau}{\hbar} \frac{1}{2n_{i,0} \xi^D} F.
\end{equation}
The Broglie wavelength in Eq. (\ref{eq:Zideal}) can also be rewritten to
\begin{equation}\label{eqn: broglie wavelength tilde}
  \tilde{\lambda} = \frac{\lambda}{\xi} \Rightarrow \tilde{\lambda} = \sqrt{\frac{2\pi}{2 n_{i,0} \xi^D}\tilde{\beta}}.
\end{equation}
To explicitly calculate the units, the term $2n_{i,0}\xi^D$ has to be filled in, this is however different for 3D and 1D quantum droplets. In section \ref{sec: results} the units will also be chosen differently for 3D and 1D droplets, so the units for 3D and 1D droplets will be treated separably.

\subsection{3D units}
The term $2n_{i,0}\xi^3$ in three dimensions can be simplified with Eqs. (\ref{eqn: infinite density}) and (\ref{eqn units xi and tau}) to
\begin{equation}\label{eqn: nixi}
    2n_{i,0}\xi^3 = \frac{96 \sqrt{6}}{5 \pi^2} \qty(\frac{g}{\abs{\delta g}})^{5/2}.
\end{equation}
This means that the free energy and transition temperature can be expressed in units of $\hbar/\tau$ and $\tau/\hbar$, and that the result is only dependent on $\abs{\delta g}/g$, $N$ and $V$. However, because the results will be compared with other works that use as their characteristic scale for the energy
\begin{equation}
    E_0=\frac{\hbar^2}{ma^2} = \frac{16\pi^2\hbar^6}{m^3}\frac{1}{g^2},
\end{equation}
the units will be further rewritten. $\tau$ can be written in terms of $E_0$ by using Eqs. (\ref{eqn units xi and tau}) and (\ref{eqn: infinite density}) to
\begin{equation}
    \tau = \frac{24576}{25 \pi^2}\qty(\frac{g}{\abs{\delta g}})^3 \frac{\hbar}{E_0}.
\end{equation}
What can be used together with Eq. (\ref{eqn: nixi}) to rewrite the prefactors for $\beta$, $E$ and $F$ to
\begin{equation}\label{eqn: 3D voorfactor}
    \frac{\hbar}{\tau} (2n_{i,0} \xi^3) =  \frac{5}{128} \sqrt{\frac{3}{2}\frac{\abs{\delta g}}{g}}E_0.
\end{equation}
Which makes it possible to write the inverse temperature and the free energy in units of $E_0$ 
\begin{equation}
    \tilde{\beta} = \qty(\frac{5}{128} \sqrt{\frac{3}{2}\frac{\abs{\delta g}}{g}}) E_0 \beta \qq{and} \tilde{F} = \qty(\frac{5}{128} \sqrt{\frac{3}{2}\frac{\abs{\delta g}}{g}})^{-1} \frac{F}{E_0}.
\end{equation}
Since $\xi$ in Eq. (\ref{eqn units xi and tau}) is dependent on $(\delta g)/g$, lengths will be expressed in terms of the scattering length ($a$) instead of $\xi$. The scattering length can be transformed in $\xi$ with the following formula
\begin{equation}
    \xi = \frac{2^{13/2}\sqrt{3}}{5\pi} \qty(\frac{g}{\abs{\delta g}})^{3/2} a.    
\end{equation}

\subsection{1D units}
Using Eq. (\ref{eqn: 1d untis}),  $2n_{i,0}\xi$ can be simplified to
\begin{equation}
    2n_{i,0}\xi = \frac{8\sqrt{2}}{9\pi} \qty(\frac{g}{\delta g})^{3/2}.
\end{equation}
However most literature express the energy in terms of the binding energy of a interspecies dimer:
\begin{equation}
    E_B = -\frac{\hbar^2}{ma_{12}^2} = -\frac{mg_{12}^2}{4\hbar^2},
\end{equation}
where Eq. (\ref{eqn gij 1d}) was used. Rewriting the prefactors for $\beta$, $E$ and $F$ with Eq. (\ref{eqn: 1d untis}) results in
\begin{equation}\label{eqn units 1D}
    \frac{\hbar}{\tau}(2n_{i,0}\xi)= \frac{16\sqrt{2}m}{9\pi^3\hbar^2}\frac{g^{9/2}}{(\delta g)^{5/2}} 
    =\frac{64\sqrt{2}}{9\pi^3}\qty(\frac{g}{\delta g})^{5/2}\qty(1-\frac{\delta g}{g})^{-2} \abs{E_B}.
\end{equation}
Now the inverse temperature and free energy can be expressed in units of $\abs{E_B}$
\begin{equation}
    \tilde{\beta} = \qty[\frac{64\sqrt{2}}{9\pi^3}\qty(\frac{g}{\delta g})^{5/2}\qty(1-\frac{\delta g}{g})^{-2}] \abs{E_B} \beta 
\end{equation}
and
\begin{equation}
    \tilde{F} = \qty[\frac{64\sqrt{2}}{9\pi^3}\qty(\frac{g}{\delta g})^{5/2}\qty(1-\frac{\delta g}{g})^{-2}]^{-1}  \frac{F}{\abs{E_B}}.
\end{equation}
Similar to the 3D case, $\xi$ is dependent on $(\delta g)/g$. so instead, the scattering length will be used as length unit. The scattering length can be transformed in $\xi$ with the following formula
\begin{equation}\label{eq lengte 1D}
    \xi = \frac{\pi\hbar^2\sqrt{\delta g}}{\sqrt{2}m g^{3/2}} = \frac{\pi\hbar^2}{\sqrt{2}m}\sqrt{\frac{\delta g}{g}} \frac{\abs{a}m}{2\hbar^2}=\frac{\pi}{2^{3/2}}\sqrt{\frac{\delta g}{g}} \abs{a}.
\end{equation}




\end{appendices}


\bibliography{sn-bibliography}

\end{document}